Production of $^{26}$Al in stellar hydrogen-burning environments: spectroscopic properties of states in $^{27}$Si

A. Parikh[1,2,3,4,*], K. Wimmer[3,4,5,†], T. Faestermann[3,4], R. Hertenberger[4,6], H.-F. Wirth[4,6], A. A. Chen[7,8], J. A. Clark[9], C. M. Deibel[9,10,‡], C. Herlitzius[3,4], R. Krücken[3,4,§], D. Seiler[3,4], K. Setoodehnia[7], K. Straub[3,4,¶], C. Wrede[11,†]

[1] Departament de Física i Enginyeria Nuclear, Universitat Politècnica de Catalunya, E-08036 Barcelona, Spain

[2] Institut d'Estudis Espacials de Catalunya, E-08034 Barcelona, Spain

[3] Physik Department E12, Technische Universität München, D-85748 Garching, Germany

[4] Maier-Leibnitz-Laboratorium der Münchner Universitäten (MLL), D-85748 Garching, Germany

[5] National Superconducting Cyclotron Laboratory, Michigan State University, East Lansing, MI 48824, USA

[6] Fakultät für Physik, Ludwig-Maximilians-Universität München, D-85748 Garching, Germany

[7] Department of Physics and Astronomy, McMaster University, Hamilton, ON L8S 4M1, Canada

[8] Excellence Cluster Universe, Technische Universität München, D-85748 Garching, Germany

[9] Physics Division, Argonne National Laboratory, Argonne, IL 60439, USA

---

[*] email: anuj.r.parikh@upc.edu

[†] Present address: National Superconducting Cyclotron Laboratory, Michigan State University, East Lansing, MI 48824, USA

[‡] Present address: Department of Physics and Astronomy, Louisiana State University, Baton Rouge, LA 70803, USA

[§] Present address: TRIUMF, Vancouver, BC, V6T 2A3, Canada

[¶] Present address: Istituto Nazionale di Fisica Nucleare, I-56127 Pisa, Italy

[10] Joint Institute for Nuclear Astrophysics, Michigan State University, East Lansing, MI 48824, USA

[11] Department of Physics, University of Washington, Seattle, WA 98195, USA

ABSTRACT

Model predictions of the amount of the radioisotope $^{26}$Al produced in hydrogen-burning environments require reliable estimates of the thermonuclear rates for the $^{26g}$Al(p,γ)$^{27}$Si and $^{26m}$Al(p,γ)$^{27}$Si reactions. These rates depend upon the spectroscopic properties of states in $^{27}$Si within about 1 MeV of the $^{26g}$Al+p threshold ($S_p$ = 7463 keV). We have studied the $^{28}$Si($^{3}$He,α)$^{27}$Si reaction at 25 MeV using a high-resolution quadrupole-dipole-dipole-dipole magnetic spectrograph. For the first time with a transfer reaction, we have constrained $J^\pi$ values for states in $^{27}$Si over $E_x$ = 7.0 – 8.1 MeV through angular distribution measurements. Aside from a few important cases, we generally confirm the energies and spin-parity assignments reported in a recent γ-ray spectroscopy study. The magnitudes of neutron spectroscopic factors determined from shell-model calculations are in reasonable agreement with our experimental values extracted using this reaction.

PACS: 21.10.Hw, 26.20.Cd, 25.55.Hp, 27.30.+t

I.   INTRODUCTION

Radioactive nuclei produced in various astrophysical phenomena may β-decay to daughter nuclei in excited states, which subsequently de-excite through the emission of characteristic γ-rays. The high penetrating power of γ-rays permits direct translation of these observables into abundances of the mother nuclei, which can then be used to constrain and test nucleosynthesis predictions from stellar models. Obtaining absolute abundances using measurements from elsewhere in the electromagnetic spectrum generally requires additional, possibly speculative assumptions regarding the environments (e.g., stellar atmospheres) under consideration; moreover, such observations generally only provide elemental (as opposed to isotopic) abundances. Since the 1.809 MeV β-delayed γ-ray line from the decay of the ground-state of $^{26}$Al ($t_{1/2}$ = 7.2 × 10$^5$ y, J$^π$ = 5$^+$) is the most thoroughly examined case [1-7], its intensity and distribution within the Galaxy provides one of the most robust constraints on nucleosynthesis predictions from theoretical models. Reproducing the inferred abundance of $^{26}$Al in the Galaxy (2.7 ± 0.7 M$_⊙$ [7]) with a single model (or several models accounting for different nucleosynthesis sites) could have far-reaching consequences. For example, $^{26}$Al is inferred to have been present in the early solar system to the level of $^{26}$Al/$^{27}$Al ~ 5 × 10$^{-5}$ (from meteoritic inclusions, see e.g., [8]) and the energy released by its decay was partially responsible for the melting and differentiation of planetesimals (see e.g., [9,10]), the first large bodies to form in the solar system. Planetesimals may in turn have been the source of much of Earth's water [11, 12], so the habitability of our planet could be directly linked to the stellar nucleosynthesis of $^{26}$Al.

Radioisotopes such as $^{26}$Al and $^{60}$Fe are long-lived relative to the recurrence timescales of events which create these isotopes. Complications in the interpretation of the γ-ray observations therefore arise because the detected intensities likely consist of the superposition of the emission from nuclei produced in different events, distributed in both time and space. For example, the 3D spatial distribution of the sources must be determined to explain the observed flux. Complications in the relevant nuclear physics behind the net production of $^{26}$Al arise from the presence of an isomeric state in $^{26}$Al at $E_x$= 228 keV ($t_{1/2}$ = 6.3 s, J$^π$= 0$^+$). (Hereafter, the ground state of $^{26}$Al will be denoted as $^{26g}$Al, the isomeric state as $^{26m}$Al, and the general nucleus as simply $^{26}$Al.) $^{26g}$Al β-decays to the $E_x$ = 1.809 and 2.938 MeV states in $^{26}$Mg, leading to a 1.809 MeV γ-ray in 99.7% of all decays. The isomer, on the other hand, decays directly (100%) to the ground-state of $^{26}$Mg without the emission of any γ-ray. Hence, one must distinguish between production and destruction of both $^{26g}$Al and $^{26m}$Al in astrophysical phenomena, particularly at temperatures T < 0.4 GK where thermal equilibrium between the ground and isomeric states is not assured [13 – 16].

Stellar winds from Asymptotic Giant Branch (AGB) and Wolf-Rayet (WR) stars, and ejection through core-collapse supernova and classical nova explosions have been suggested as mechanisms through which $^{26}$Al may be distributed throughout the interstellar medium [4, 6, 16 – 22]. Each of these scenarios involves different characteristic temperatures over which $^{26}$Al is most likely to be produced or destroyed. As a result, nuclear structure information for different nuclei, and over various excitation energies in these nuclei, is required to characterize all of the relevant thermonuclear reaction rates.

In hydrogen-burning environments, the $^{26}$Al(p,γ)$^{27}$Si reaction is the key pathway for the destruction of $^{26}$Al. The winds of AGB and WR stars are thought to eject $^{26}$Al produced through hydrogen-burning at temperatures between roughly 30 – 100 MK. In classical novae, $^{26}$Al is produced only in explosions that involve an oxygen-neon white dwarf and achieve the highest temperatures, e.g., $T_{peak} \approx$ 0.2 – 0.4 GK. These considerations imply the need to understand the nuclear structure of $^{27}$Si in the energy range between the $^{26g}$Al+p threshold ($S_p$ = 7462.96(16) keV [23]) and roughly 1 MeV above this threshold to reliably evaluate the $^{26g}$Al(p,γ)$^{27}$Si and $^{26m}$Al(p,γ)$^{27}$Si rates in all of these environments. In particular, one requires the resonance energies $E_R$ and (p,γ) resonance strengths for states in this energy range. Unknown resonance strengths may be estimated using indirect techniques, for example, using proton-transfer spectroscopic factors, proton and γ-decay branching ratios and $J^π$ values – see e.g., Iliadis (2007) [24] for more details.

The $^{26}$Al(p,γ)$^{27}$Si reaction has been studied both directly and indirectly. Buchmann et al. (1984) [25] measured an excitation function using protons bombarding a $^{26g}$Al target, and found 7 resonances within $E_R$ = 270 - 900 keV ($E_x$ = 7.74– 8.36 MeV). They also used the thick-target method to find the strengths of these resonances. Schmalbrock et al. (1986) [26] used the $^{28}$Si($^3$He,α)$^{27}$Si reaction to determine the energies of 58 states from $E_x$ = 4.14 – 8.37 MeV; 18 of these states had excitation energies greater than 7.46 MeV. These measurements were largely confirmed by Wang et al. (1989) [27], who studied both the $^{27}$Al($^3$He,t)$^{27}$Si and $^{28}$Si($^3$He,α)$^{27}$Si reactions. Both Schmalbrock et al. [26] and Wang et al. [27] attempted unsuccessfully to determine $J^\pi$ values for proton-threshold states in $^{27}$Si through the measurement of angular distributions for the $^{28}$Si($^3$He,α)$^{27}$Si reaction. Contaminant alpha groups from ($^3$He,α) reactions on carbon and oxygen were an issue in both studies; Wang et al. [27] nonetheless extracted angular distributions, but did not attempt to fit them. Vogelaar et al. (1996) [28] measured the $^{26g}$Al($^3$He,d)$^{27}$Si reaction to constrain strengths for states at $E_x$ = 7.59, 7.65, and 7.74 MeV through estimates of the respective proton-transfer spectroscopic factors. Ruiz et al. (2006) [29], through a direct study in inverse kinematics with a high-intensity beam of $^{26g}$Al, measured both the energy and strength of the key resonance for $^{26g}$Al destruction in classical nova explosions ($E_R$ = 184 keV, $E_x$ = 7.65 MeV). Both the resonance energy and strength, however, were in minor disagreement with the previous unpublished direct study of Vogelaar (1989) [30] (but see section IV), and with the results from the transfer reaction studies of [26] and [27]. Deibel et al. (2009) [31] observed 53 states between $E_x$ = 8.14 and 9.86 MeV through studies of the $^{27}$Al($^3$He,t)$^{27}$Si and $^{28}$Si($^3$He,α)$^{27}$Si reactions. They also observed proton decays from excited states in $^{27}$Si between $E_x$ = 8.14 – 8.98 MeV to the ground-, isomeric and second-excited states of $^{26}$Al, and so were able to constrain the corresponding proton-branching ratios.

Outstanding nuclear physics issues for $^{26}$Al destruction in AGB and WR stars and classical nova explosions included (a) $^{26g}$Al(p,γ) resonance strengths (or spectroscopic information) for states $E_x$($^{27}$Si) < 7.65 MeV; (b) the possible existence of the state at $E_x$($^{27}$Si) = 7.56 MeV (identified only in the studies of Wang et al. (1989) [27]); (c) the energy of the resonance at $E_x$($^{27}$Si) = 7.65 MeV, given the ≈4 keV disagreement between the Ruiz et al. [29] study and the other studies [26, 27, 30]. Points (a) and (b) are of particular relevance for $^{26}$Al production in AGB and WR stars as they gave rise to uncertainties of ≈4 orders of magnitude in the $^{26g}$Al(p,γ)$^{27}$Si destruction rate over the temperatures involved [32], which, in turn, could affect $^{26}$Al production in e.g., AGB stars by factors of ≈100 [33, 34]. Point (c) is of importance to precisely quantify $^{26}$Al production in classical novae (expected to contribute less than 20% of the overall Galactic $^{26}$Al abundance [22, 35]). The contribution to a thermonuclear rate of a narrow, isolated resonance depends linearly on the strength of the resonance and exponentially on the resonance energy $E_R$ through a factor exp(-$E_R$/kT), where T is the temperature of interest and k is the Boltzmann constant. An uncertainty of 4 keV in $E_R$ leads to an uncertainty of ≈20% in the contribution of this resonance to the overall $^{26g}$Al(p,γ)$^{27}$Si rate at typical nova peak temperatures (over which this resonance dominates the reaction rate). This uncertainty is comparable to the uncertainties in measurements of the strength of this resonance [29, 30]. Moreover, a reduction in the $^{26g}$Al(p,γ)$^{27}$Si rate by ≈20% led to an increase in the overall yield of $^{26}$Al by ≈20% in the nova model discussed in [29]. Finally, in addition to the above, a reliable $^{26m}$Al(p,γ)$^{27}$Si rate would require resonance strengths (or complete spectroscopic information) for states between $E_x$ = 7.69 MeV (i.e., the $^{26m}$Al+p energy threshold in $^{27}$Si) and at least $E_x$ = 8.14 MeV (above which proton branching ratios to the ground and isomeric states have been measured [31]).

Lotay et al. [36] addressed some of these issues through a detailed γ-ray spectroscopy study of $^{27}$Si using the fusion-evaporation reaction $^{12}$C($^{16}$O,n). $J^π$ values were assigned for levels between the ground state and $E_x$ = 8.38 MeV. No state at 7.56 MeV was observed, and the excitation energy of the 7.65 MeV state was found to be in disagreement with the value from [29]. With this new information, a $^{26g}$Al(p,γ)$^{27}$Si rate was recently determined with uncertainties of less than a factor 10 below 0.1 GK, and less than ≈20% above 0.1 GK [37]. A reliable $^{26m}$Al(p,γ)$^{27}$Si rate still cannot be calculated, however, due to insufficient experimental information [16].

The structure of $^{27}$Si within about 1 MeV of the $^{26g}$Al+p threshold is of clear importance in constraining the $^{26g}$Al(p,γ) and $^{26m}$Al(p,γ) rates in hydrogen-burning environments such as AGB and WR stars and classical nova explosions. Given this, we have performed a high-resolution study of the $^{28}$Si($^{3}$He,α)$^{27}$Si reaction to independently determine the energies and $J^π$ values for relevant states in $^{27}$Si. We also desired to test the assertion of Wang et al. (1989) [27] that reliable $J^π$ values could not be extracted from angular distributions of the $^{28}$Si($^{3}$He,α)$^{27}$Si reaction at E ≈ 22 MeV, for states above the $^{26g}$Al+p energy threshold in $^{27}$Si.

## II. EXPERIMENT

The $^{28}$Si($^{3}$He,α)$^{27}$Si reaction was measured at the Maier-Leibnitz-Laboratorium (MLL) in Garching, Germany over a total period of three days. A 25 MeV beam of $^{3}$He$^{2+}$ ions (I ≈ 550 nA) was produced with an electron-cyclotron-resonance-like ion source [38] and an MP tandem accelerator. This beam was transported to the target position of a quadrupole-dipole-dipole-dipole (Q3D) magnetic spectrograph with superior intrinsic energy resolution $\Delta E/E \approx 2 \times 10^{-4}$ [39]. Targets were prepared at the Technische Universität München, and included: enriched silicon (20 μg/cm$^2$, enriched to 99.84% $^{28}$Si) deposited upon a foil of enriched carbon (7 μg/cm$^2$, enriched to 99.99% $^{12}$C); natural silicon dioxide (self-supporting, 25 μg/cm$^2$); enriched carbon (10 μg/cm$^2$ foil, enriched to 99.99% $^{12}$C); and enriched magnesium (20 μg/cm$^2$, enriched to 99.92 % $^{24}$Mg) deposited upon a foil of enriched carbon (7 μg/cm$^2$, enriched to 99.99 % $^{12}$C). The carbon and silicon dioxide targets were used primarily to characterize background due to reactions on the carbon and oxygen present in the enriched silicon target, and the magnesium target was used to help calibrate the focal-plane of the spectrograph. Light reaction products entered the Q3D spectrograph through a rectangular aperture (encompassing 7.0 msr), were dispersed according to their momenta, and finally, were focused onto a multi-wire gas-filled proportional counter backed by a plastic scintillator [40]. Alpha particles were clearly identified through energy loss and residual energy information from the focal-plane detection system, and alpha spectra of focal-plane position were then produced. Measurements were made at spectrograph angles of 10, 15, 20, 25, 30, 35, 40, 45, 55, and 65°; the beam current was integrated using a Faraday cup placed at 0° in the target chamber.

III. DATA AND ANALYSIS

Fig. 1 shows alpha spectra measured with the enriched silicon target at spectrograph angles of 15° and 20°. Contaminant groups due to ($^3$He,α) reactions on $^{12}$C and $^{16}$O present in the target (the former primarily from the target backing) were evident, and these were unambiguously identified and characterized through both kinematic analysis at the measured angles and measurements with the enriched carbon and silicon dioxide targets. For example, contaminant groups due to $^{16}$O($^3$He,α) reactions populating the 8743(6), 8922(2) and 8982.1(17) keV states in $^{15}$O [41] are seen among the 7380, 7534, 7592 and 7652 keV states of $^{27}$Si in Fig. 1a and among the 7534, 7652, 7694, 7704 and 7740 keV states of $^{27}$Si in Fig. 1b. These spectra were analyzed using least-squares fits of multiple Gaussian or exponentially-modified Gaussian functions. Consistent excitation energies were determined using each of these prescriptions. Peak widths were fixed to ≈12 keV FWHM based on fits of isolated peaks in these spectra.

At each measurement angle the focal-plane was calibrated using well-resolved, known states in $^{23}$Mg (7.6 < $E_x$($^{23}$Mg) < 8.7 MeV and Δ$E_x$ = 1 – 6 keV [42, 43]) populated via the $^{24}$Mg($^3$He,α) reaction with the enriched magnesium target. With this information, second-degree polynomial fits of alpha radius of curvature ρ to focal-plane position yielded excitation energies for states in $^{27}$Si. Excitation energies from the present work are listed in Table I, along with uncertainties due to counting statistics, reproducibility among angles, and uncertainties in the calibration states; the slightly larger uncertainties for states with $E_x$ > 7.9 MeV arise due to the increasing reliance on calibration states with larger uncertainties. The energies from the present work are all weighted averages calculated with energies determined for at least four different angles – the exact number depended upon the precise magnetic field setting used at a particular angle (i.e., for states near the edges of the focal-plane), the presence of large contaminant peaks obscuring different states at different angles, and the requirement that a $^{27}$Si state lie within a region spanned entirely by calibration peaks. As well, we note a systematic uncertainty of ± 2 keV due to uncertainty in the thicknesses of the enriched silicon and enriched magnesium targets (each target thickness is known to roughly 10%) and uncertainty in the relative Q-value of the $^{28}$Si($^{3}$He,α)$^{27}$Si and $^{24}$Mg($^{3}$He,α)$^{23}$Mg reactions (this last aspect is dominated by the 0.8 keV uncertainty in the mass of $^{23}$Mg [23, 44]).

Angular distributions measured using the $^{28}$Si($^{3}$He,α)$^{27}$Si reaction are plotted in Fig. 2, along with direct reaction calculations using the code FRESCO [45]. Only well-resolved singlet states clearly observed over at least five angles are included in Fig. 2. Optical model parameters for the calculations were obtained using global scaling formulas [46] for the incoming $^{28}$Si + $^{3}$He channel. For the outgoing channel, parameters were taken from a study of the elastic scattering of alpha particles on the isobar $^{27}$Al at the same incident energy [47]. The shapes of the corresponding angular distribution calculations were found to be insensitive to modest variations of these optical model parameters. A further improvement was made through the consideration of inelastic excitations to the $2_1^+$ state in $^{28}$Si. This improves the agreement between the calculations and the experimental data at large angles and generally reduces the amplitudes of the oscillations in the differential cross sections, but does not influence the extracted angular momentum transfers L. As in all one-particle transfer reactions, the shape of the angular distribution is insensitive to the spin J of the final state, and so only the angular momentum transfer L can be determined. This allows one to determine the parity of states populated in the reaction, making our results completely complementary to those from the γ-ray spectroscopy measurement of [36]. In that measurement, γ-ray branching ratios and angular distributions were used to determine the ΔI of γ-ray transitions, which were then used to assign spins to excited states of $^{27}$Si; the corresponding parities of the states were inferred largely (but not exclusively) through comparisons with states in the mirror nucleus $^{27}$Al.

IV. DISCUSSION

In Table I we compare results from the present study to previous studies of the structure of $^{27}$Si between $E_x$ = 7.0 – 8.1 MeV. We include in Table I energies for peaks observed in the present work that may coincide with previously-identified doublets. For example, the peak at 7.074 MeV in the present work was observed with a somewhat larger width than other, isolated states, and, falls between the 7.059 and 7.080 MeV states determined in the study of [26]. Similar considerations apply for the 7.334, 7.434 and 7.699 MeV peaks observed in the present work; note, however that the 7.074 and 7.434 MeV peaks coincide with single states observed in [36]. Attempts to analyze these peaks as unresolved doublets did not produce significant improvements over single-level fits. As well, no appreciable changes in the shapes of these peaks as a function of angle were observed.

Energies determined in the present study generally agree well with values from previous measurements. The observation of states at 7.25 MeV [36] and 7.26 MeV (present work and [26]) indicate the existence of a previously-unidentified doublet. We do not observe a state at 7.49 MeV nor at 7.56 MeV (tentatively identified in the studies of [36] and [27], respectively). For the state at 7.65 MeV, our energy is in agreement with the values from the previous transfer-reaction and γ-ray spectroscopy measurements [26, 27, 36]. The discrepancy between this energy and that from the radiative proton-capture measurement of [29] could possibly be explained by the excitation of different members of a closely-spaced doublet in $^{27}$Si by the different experiments. Support for this hypothesis comes from the different γ-decay schemes reported for this state by [36] and in the unpublished proton-capture measurement of Vogelaar [30] – see [48] for more details. As well, the resonance energy determined in [30] ($E_p$ = 195.6(11) keV, $E_x$ = 7651.3(11) keV) may need to be reduced by a few keV due to an adjustment in the energy of a calibration state; this would improve the accord between the energies determined from the two proton-capture studies [29, 30]. Measurements are in progress to explore this issue further [49]. A previously-unresolved doublet of states at 7.832 and 7.838 MeV was observed for the first time by [36]; we could not resolve these two states, but we do note that our energy for this doublet (7.831 MeV) indicates the weak relative population of the higher-energy member at all angles. Finally, the presence of a previously-unknown doublet seems required by the observation of a state at 7.890(2) MeV (in the present study) and a state at 7.8990(8) MeV (in the study of [36]).

In Table I we have listed spin-parity constraints from the present work that arise directly from our measured angular distributions – we have not appealed to any tentative mirror assignments (e.g., those suggested by [36]). Except for the assignments to the states at 7.00, 7.13, 7.91, 8.03 and 8.07 MeV, all of our constraints are compatible with the $J^\pi$ values assigned in [36]. Calculated angular distributions for L values corresponding to the best-fit cases as well as those values deduced from [36] are plotted in Fig. 2 for the three states above the $^{26g}Al+p$ threshold; the experimental data at low angles in particular favours our $J^\pi$ constraints. The disagreement in assigned parity for the 7.00 and 8.07 MeV states may arise from incorrect mirror assignments in [36].

Measured angular distributions from the $^{28}Si(^3He,\alpha)^{27}Si$ reaction have been published previously for the 7.47, 7.53, 7.59 and 7.65 MeV states by Wang et al. [27]. Although a similar beam energy was employed in that study, the angular range was limited to $\theta_{cm}$ less than $\approx 35°$ for these four states. The authors expressed the lack of easily discernible direct reaction characteristics for their angular distributions and thus did not extract spin and parity information. In contrast, our measured angular distributions show characteristic features mainly because the angular range has been significantly extended (up to $\theta_{cm} = 72°$ for these states). The relative cross-sections agree quite well between the present study and that of [27], although absolute values differ by a factor of $\approx 3$ for common angles. The source of this discrepancy is not understood – the slight increase in beam energy (22.4 MeV in Wang et al. [27], versus 25 MeV in the present study) does not account for such differences. The general good agreement between our $J^\pi$ constraints and those from the γ-ray study of [36] indicate that treating the neutron-removal $^{28}$Si($^3$He,α) process as a direct reaction at these energies is reasonable.

Assignments between analogue states in $^{27}$Si and $^{27}$Al have been extensively discussed in Lotay et al. [36], and were indeed exploited by necessity to extract many of their adopted parities for states in $^{27}$Si. To facilitate the comparison between observed states in $^{27}$Si and shell model calculations, we have determined experimental neutron-removal spectroscopic factors S for the population of the $^{27}$Si states shown in Fig. 2, where S is the ratio between the experimental and calculated differential cross-section for a state. (Note that these neutron-removal spectroscopic factors from the $^{28}$Si($^3$He,α) reaction are not of interest for calculations of the thermonuclear $^{26}$Al(p,γ)$^{27}$Si rate. For such applications we would require proton-transfer spectroscopic factors, which could be determined through measurement of e.g., the $^{26}$Al($^3$He,d)$^{27}$Si reaction – see [28].) Theoretical energy levels and neutron spectroscopic factors were calculated in the shell-model using the code OXBASH [50]. Within the model space of the USDA interaction [51] only states with $J^\pi$ = {1/2$^+$, 3/2$^+$, 5/2$^+$} are expected to be populated in the one-neutron removal reaction. In Fig. 3 we compare the experimental spectroscopic factors for these low-spin states to the calculated values; the experimental values are also tabulated in Table II. Over $E_x$($^{27}$Si) = 6.9 – 8.3 MeV, the calculated spectroscopic factors for these levels vary between about 0.001 and 0.01, as may be expected for such high excitation energies. Given the discussion above on the disagreement between the absolute cross-sections of the present measurement and those of Wang et al. [27], one should consider a systematic uncertainty of a factor ≈3 in the experimental values of S. As well, for states without a definite spin-parity assignment, or where our $J^\pi$ constraints disagree with the assignments of Lotay et al. [36] (see Table 1), we have adopted the calculations with the higher spins when extracting the experimental spectroscopic factors. This affects the experimental value of S by e.g., 30% when comparing S values determined with $J^\pi = 3/2^+$ versus $J^\pi = 5/2^+$ for the state at $E_x$ = 7134 keV. From Fig. 3, we see that although the general agreement between the magnitudes of the experimental and calculated spectroscopic factors is acceptable, the direct assignment of experimental states to shell-model states is not straightforward. This is due to the high density of observed states as well as the fact that fewer low-spin states are predicted in this energy region than have been observed (see Table I – only states that were well-resolved in the present set of measurements are included in Fig. 3). We note, however, that significant progress has been made in this regard by Lotay et al. [36] through concurrent γ-ray spectroscopy studies of $^{27}$Al and $^{27}$Si.

The structure of $^{27}$Si above the $^{26g}$A+p threshold (7463 keV) and the $^{26m}$Al+p threshold (7691 keV) is required to calculate thermonuclear rates for proton-capture on the ground and isomeric states of $^{26}$Al. We do not observe a state at 7.56 MeV, and the energy we determine for a state at 7.65 MeV is consistent with the measurements of [26, 27, 36] (and inconsistent with the energy from the study of [29]). Comparison between the constraints of the present study and [36] indicate that the 7.47 MeV state has $J^\pi = 5/2^+$. Given these considerations, as well as our agreement with the $J^\pi$ values of [36] for the 7.53 and 7.59 MeV states, we propose no changes to the thermonuclear $^{26g}$Al(p,γ)$^{27}$Si rate determined in the recent evaluation of [37] over temperatures relevant to hydrogen-burning in AGB and WR stars and classical nova explosions. It is still not possible to calculate a reliable experimental $^{26m}$Al(p,γ)$^{27}$Si rate [16] because of the lack of resonance strength (or proton spectroscopic factor) measurements for states immediately above the $^{26m}$Al+p threshold. The spin-parities of these states as determined in the present work and that of [36] therefore represent critical information needed both for rate estimates and to guide future experimental investigations dedicated to improving the $^{26m}$Al(p,γ) rate. Indeed, Lotay et al. [36] express the importance of the 8.07 MeV state given their $J^\pi$ assignment of $3/2^-$ (corresponding to an l = 1 proton-capture resonance). Our data, however, is consistent with a different assignment for both the 8.07 MeV state (i.e., corresponding to an l = 2 resonance) and the 7.91 MeV state (i.e., corresponding to an l = 1 resonance). This would shift the importance of the l = 1 resonance down to lower temperatures, where it could dominate the reaction rate [36]. Without more nuclear structure information however, it is impossible at the moment to precisely quantify the contributions of individual resonances to the $^{26m}$Al(p,γ) rate at temperatures involved in AGB and WR stars and classical nova explosions. The existence of additional states beyond those in Table I should also be investigated.

## V. CONCLUSIONS

We have measured energies and angular distributions for states in $^{27}$Si over $E_x$ = 7.0 – 8.1 MeV through a study of the $^{28}$Si($^{3}$He,α) reaction. Constraints on $J^\pi$ values for sixteen states above the $^{26g}$Al+p threshold have been determined for the first time using a transfer reaction; these constraints (and all energies) are generally in good agreement with the results from a recent γ-ray spectroscopy study [36]. A direct reaction mechanism adequately describes our experimental angular distributions, in contrast with indications from a previous measurement using the same reaction and similar beam energy [27].

In the absence of measured resonance strengths for states immediately above the $^{26g}$Al+p and $^{26m}$Al+p energy thresholds, the $J^\pi$ values of these states represent critical nuclear structure information needed to estimate the corresponding thermonuclear proton-capture reaction rates. The present work confirms the assumptions made in the calculation of [37] for the $^{26g}$Al(p,γ)$^{27}$Si rate. To further reduce the uncertainties in this rate, especially over temperatures encountered in AGB and WR stars and classical nova explosions, the $^{26}$Al($^{3}$He,d)$^{27}$Si reaction should be studied to extract proton spectroscopic factors for the 7.53 and 7.59 MeV states. A vital improvement over the previous study of this reaction [28] would involve the minimization of any $^{27}$Al contamination in the target [52]. Measurements to better constrain the $^{26m}$Al(p,γ)$^{27}$Si rate could involve a study similar to that of [31], optimized to allow the detection of protons from the decay of states $E_x$($^{27}$Si) < 8.1 MeV. The reaction could also be measured directly in inverse kinematics [53]; prior identification of l = 0 and l = 1 resonances would help to guide this challenging study. For this reason, the $J^\pi$ values of states above the $^{26m}$Al+p threshold, particularly those at 7.91 and 8.07 MeV, should be confirmed.

Acknowledgments

It is a pleasure to thank the crew of the MLL tandem accelerator. We also appreciate comments from D. A. Hutcheon, J. José, A. Karakas, R. Longland, M. Lugaro and C. Ruiz. Thanks to G. Lotay and collaborators for an advance copy of their recent article. This work was supported by the DFG cluster of excellence "Origin and Structure of the Universe" (www.universe-cluster.de). AP was partially supported by the Spanish MICINN grants AYA2010-15685 and EUI2009-04167, by the E.U. FEDER funds, and by the ESF EUROCORES Program EuroGENESIS. AAC was supported, in part, by a grant from NSERC Canada. JAC and CMD acknowledge support from the U.S. Department of Energy, Office of Nuclear Physics, under Contract No. DE-AC02-06CH11357. CMD was also partially supported by JINA grant No. PHY0822648. CW acknowledges support from the U.S. Department of Energy under Contract No. DE-FG02-97ER41020.

TABLE I: Level structure of $^{27}$Si for $E_x$ = 7.0 – 8.1 MeV. Excitation energies are given in keV. Resonance energies from the studies of [25] and [29] have been converted to excitation energies using the $^{26g}$Al+p energy-threshold of $S_p$ = 7462.96(16) keV [23] in $^{27}$Si. One should consider a systematic uncertainty of ±2 keV in addition to the uncertainties listed for the present work (see text).

| $^{26}$Al(p,γ) [25] | $^{28}$Si($^3$He,α) [26] | $^{27}$Al($^3$He,t) [27] | p($^{26}$Al,γ) [29] | $^{12}$C($^{16}$O,n) [36] | $^{28}$Si($^3$He,α) Present |
|---|---|---|---|---|---|
| | 7005(8) | | | 7000.7(22) 9/2$^+$ | 7004(1) (9/2, 11/2)$^-$ |
| | 7059(5) | | | 7070.2(4) 9/2- | 7074(1) |
| | 7080(3) | | | | |
| | 7134(5) | | | 7129.0(2) 13/2$^+$ | 7134(2) (3/2, 5/2)$^+$ |
| | 7223(4) | | | 7222.4(2) 13/2$^+$ | 7225(2) (11/2, 13/2)$^+$ |
| | 7239(4) | | | 7245.4(5) 11/2$^+$ | |
| | 7260(4) | | | 7252.5(2) 7/2$^+$ | 7262(2) (5/2, 7/2)$^-$ |
| | 7276(3) | | | | |
| | 7324(4) | | | 7325.4(18) 3/2$^+$ | 7334(2) |
| | 7341(4) | | | 7346.6(9) 7/2$^-$ | |
| | 7388(5) | 7379(4) | | 7380.4(15) 5/2$^+$ | 7380(2) (3/2, 5/2)$^+$ |
| | 7436(4) | 7428(4) | | 7433.3(6) 9/2$^+$ | 7434(2) |
| | | 7436(4) | | | |
| | 7465(5) | 7470(4) | | 7468.8(8) (1/2, 5/2)$^+$ | 7472(2) (3/2, 5/2)$^+$ |
| | | | | (7493.1(40)) (3/2$^+$) | |
| | 7530(5) | 7533(3) | | 7531.3(7) 5/2$^+$ | 7534(2) (3/2, 5/2)$^+$ |
| | | (7557(3)) | | | |
| | 7596(4) | 7589(3) | | 7590.1(9) 9/2$^+$ | 7592(2) (7/2, 9/2)$^+$ |
| | 7654(5) | 7651(3) | 7647(1) | 7651.9(6) 11/2$^+$ | 7652(2) (11/2, 13/2)$^+$ |
| | | (7690(3)) | | 7693.8(9) 5/2$^+$ | 7699(2) |
| | 7703(3) | 7702(3) | | 7704.3(2) 7/2$^-$ | |
| 7738.9(3) (7/2 –11/2)$^+$ | 7742(3) | 7741(3) | | 7739.3(4) 9/2$^+$ | 7740(2) (7/2, 9/2)$^+$ |
| | 7796(4) | 7789(3) | | 7794.8(19) 7/2$^+$ | 7789(1) (7/2, 9/2)$^+$ |
| 7825(3) (7/2 –11/2)$^+$ | | 7832(3) | | 7831.5(5) 9/2$^-$ | 7831(1) |
| | 7837(4) | | | 7837.6(2) | |

| | | | | 5/2⁺ | |
| --- | --- | --- | --- | --- | --- |
| | | 7893(4) | | 7899.0(8) 5/2⁺ | 7890(2) (3/2, 5/2)⁺ |
| | 7909(4) | 7913(3) | | 7909.1(7) 3/2⁺ | 7910(3) (1/2, 3/2)⁻ |
| | 7974(5) | 7971(3) | | 7966.3(8) 5/2⁺ | 7967(3) (3/2, 5/2)⁺ |
| | 8034(5) | 8037(3) | | 8031.5(11) 5/2⁺ | 8033(3) (7/2, 9/2)⁺ |
| | 8077(5) | 8073(3) | | 8069.6(30) 3/2⁻ | 8069(3) (3/2, 5/2)⁺ |

TABLE II: Neutron spectroscopic factors S extracted from the ²⁸Si(³He,α)²⁷Si reaction, assuming the transferred angular momentum values L from the best-fit curves of Fig. 2.

| $E_x$ (²⁷Si) (keV) | L | S |
| --- | --- | --- |
| 7004 | 5 | 0.042 |
| 7134 | 2 | 0.013 |
| 7225 | 6 | 0.16 |
| 7262 | 3 | 0.028 |
| 7380 | 2 | 0.013 |
| 7472 | 2 | 0.012 |
| 7534 | 2 | 0.0079 |
| 7592 | 4 | 0.0074 |
| 7652 | 6 | 0.064 |
| 7740 | 4 | 0.0049 |
| 7789 | 4 | 0.0095 |
| 7890 | 2 | 0.012 |
| 7910 | 1 | 0.0089 |
| 7967 | 2 | 0.0020 |
| 8033 | 4 | 0.023 |
| 8069 | 2 | 0.0098 |

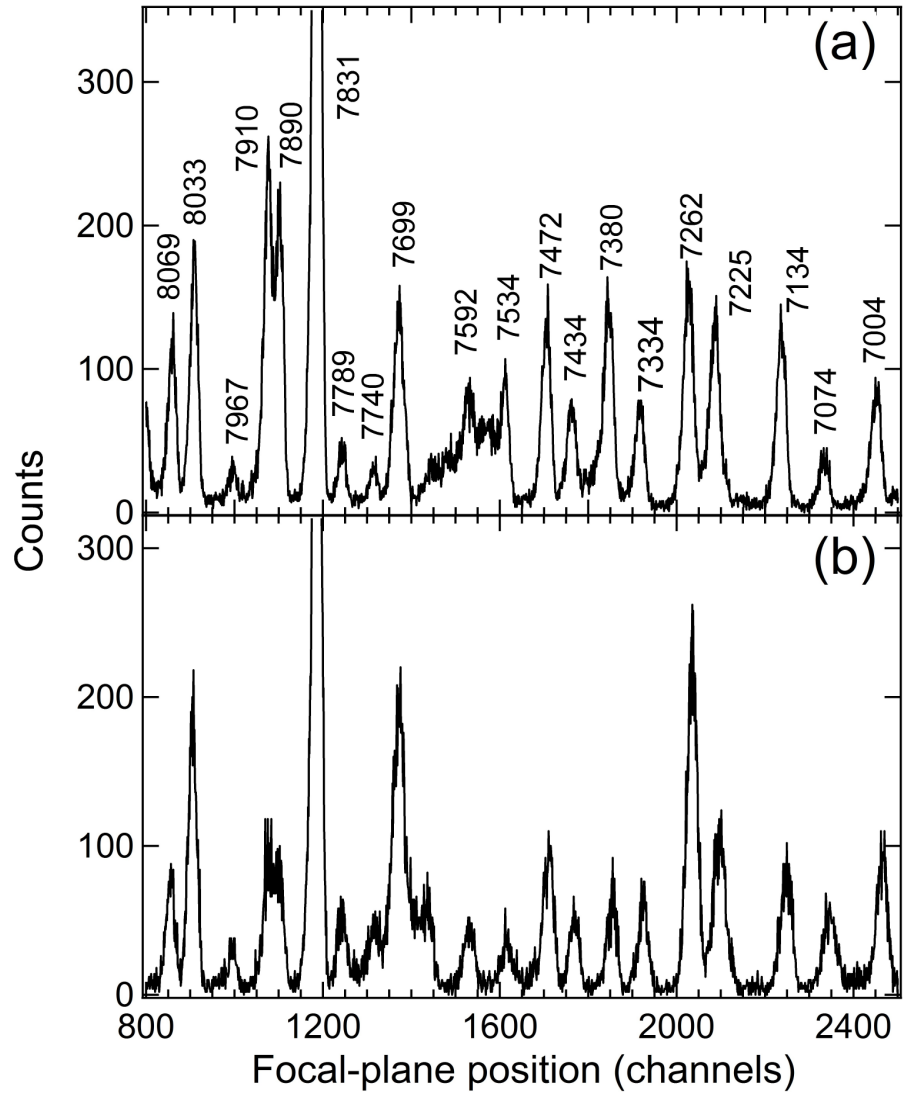

FIG. 1: Focal-plane alpha spectra from the $^{28}$Si($^3$He,α)$^{27}$Si reaction at 25 MeV, dΩ = 7.0 msr and (a) θ$_{lab}$ = 15°, (b) θ$_{lab}$ = 20°. Excitation energies are labeled in keV.

2121

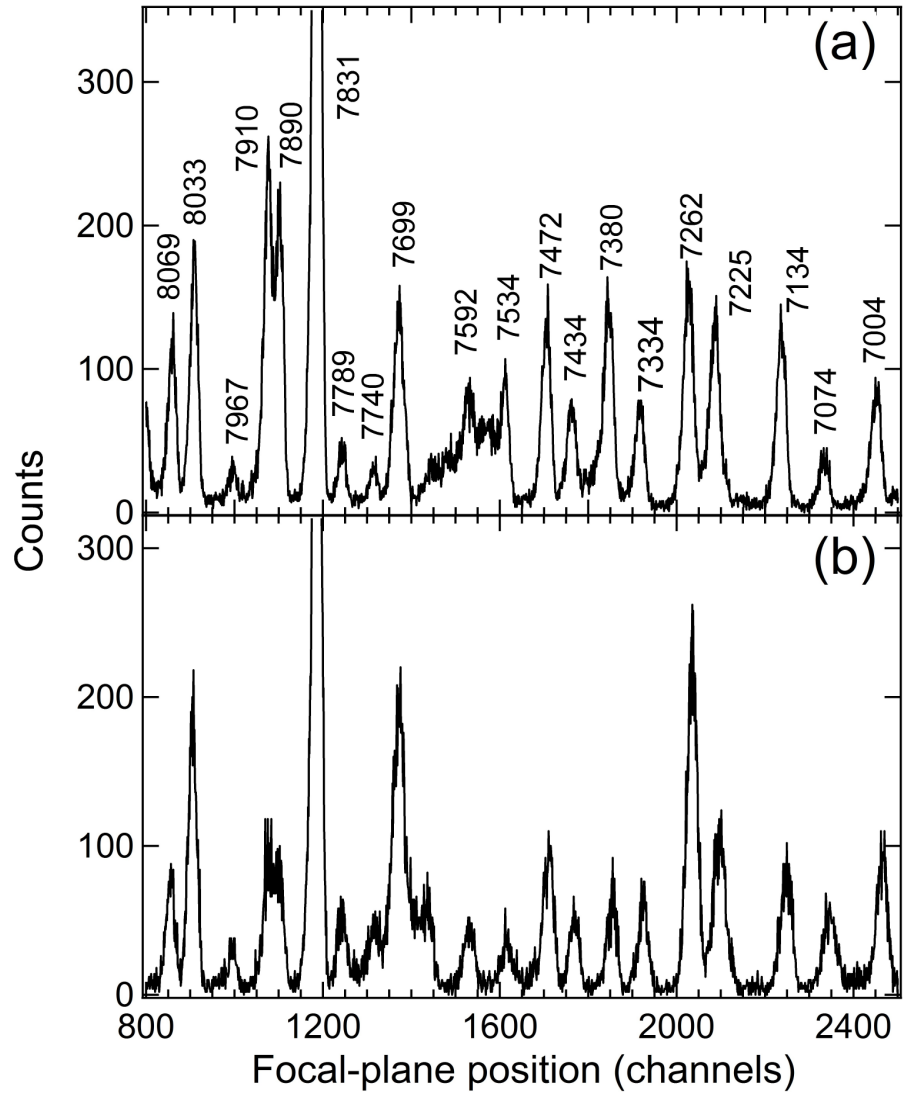

FIG. 1: Focal-plane alpha spectra from the $^{28}$Si($^3$He,α)$^{27}$Si reaction at 25 MeV, dΩ = 7.0 msr and (a) θ$_{lab}$ = 15°, (b) θ$_{lab}$ = 20°. Excitation energies are labeled in keV.

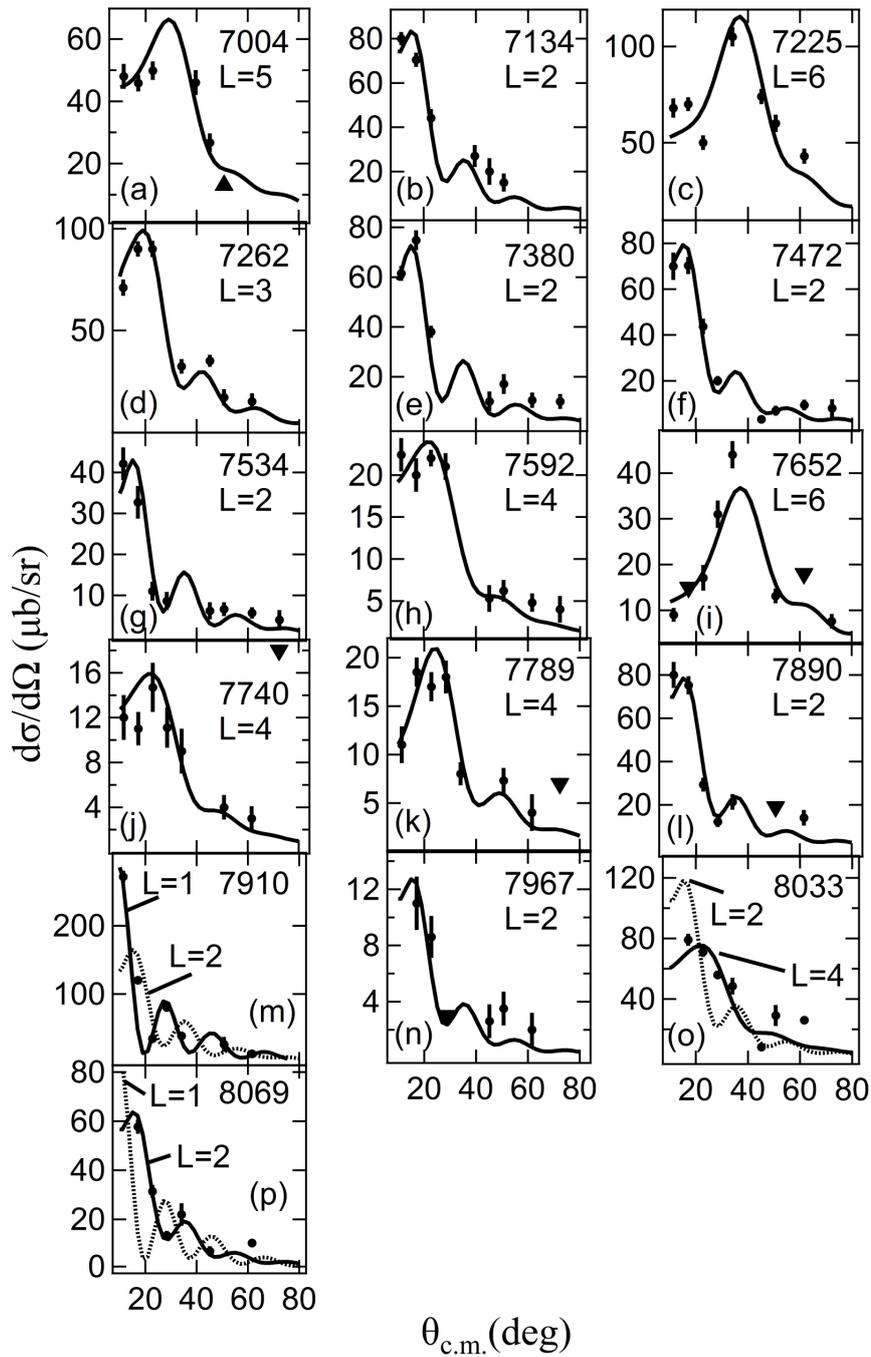

FIG 2: Alpha angular distributions measured with the $^{28}$Si($^{3}$He,α)$^{27}$Si reaction at 25 MeV. Curves calculated with the finite-range, coupled-reaction channels code FRESCO [45] have been fit to the data. Each panel (a - p) is labeled with the excitation energy (in keV) of the relevant state in $^{27}$Si and the transferred angular momentum L from the calculation that best fits the data. Panels (m), (o) and (p) also include calculations using alternative L values deduced from the spin-parity constraints of [36].

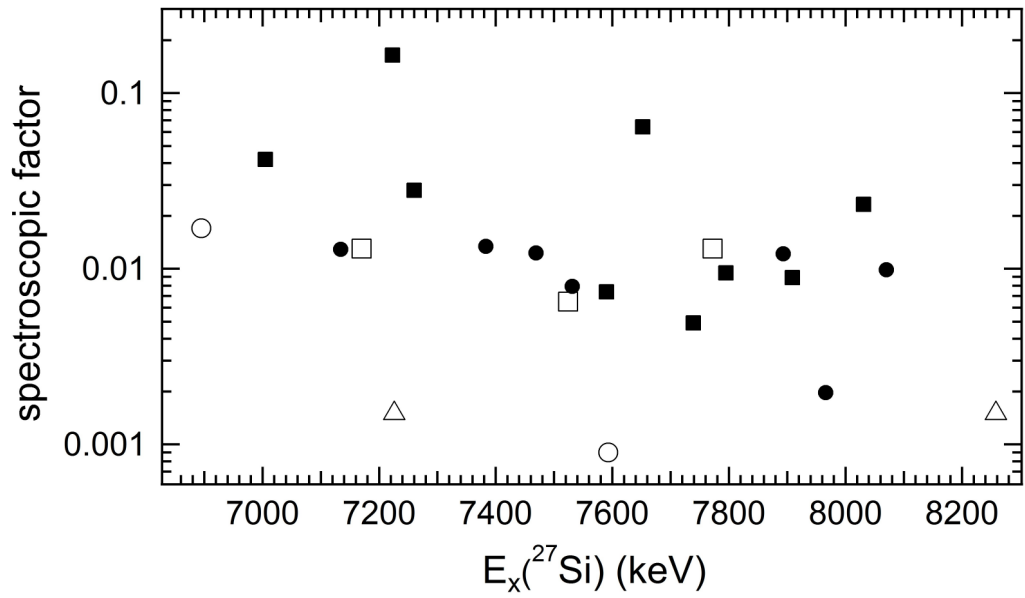

FIG. 3: Neutron spectroscopic factors for states in $^{27}$Si populated through the $^{28}$Si($^3$He,α)$^{27}$Si reaction. Experimental values for low-spin states ($J^\pi = \{1/2^+, 3/2^+, 5/2^+\}$, filled circles) and higher spin states (filled squares) are plotted with shell-model calculations for low-spin states ($J^\pi = 1/2^+$, open triangles; $J^\pi = 3/2^+$, open squares; $J^\pi = 5/2^+$, open circles).